\documentclass[11pt]{article}
\begin{document}
%%%%%%%%%%%%%% %%% Definitions:  %%%%%%%%%%%%%% 
%definition de la font pour
%R,C,N 
% \font\blackboard=msbm10 % scaled \magstep1 % \font\blackboards=msbm7
%\font\blackboardss=msbm5 
% \newfam\black \textfont\black=\blackboard 
%
%\scriptfont\black=\blackboards \scriptscriptfont\black=\blackboardss 
%
%\def\Bbb#1{{\fam\black\relax#1}} %definition de la font pour l'identite
%\font\ninerm=cmr9 %\def\uniset{\rlap{\ninerm 1}\kern.15em 1}
%\def\e{\mathop{\rm e}\nolimits} 
%Fractions 
\def\half{\scriptstyle{\frac{1}{2}}}
\def\halft{\textstyle{\frac{1}{2}}} 
\def\osqrt{\textstyle{\frac{1}{\sqrt2}}}
\def\lsqrt{\textstyle{\frac{\l}{\sqrt2}}} 
\def\phalf{\textstyle{\frac{\pi}{2}}}
\newcommand{\fscr}[2]{\scriptstyle \frac{#1}{#2}} 
%Definitions 
\def\cc{\hbox{C\kern-0.55em\raise0.4ex\hbox{$\scriptstyle |$}}}
\def\zz{\hbox{{\sf Z}\kern-0.45em\raise0.0ex\hbox{\sf Z}}}     
\def\nat{\hbox{{\sf |}\kern-0.45em\raise0.0ex\hbox{\sf Í}}}     
\def\rr{\hbox{R\kern-0.55em\raise0.4ex\hbox{$\scriptstyle |$}}}  
\def\ena{1\hskip-0.25truecm 1}  
\def\zita{{\zz}^{ 2}_{N}}
\def\zitas{{\zz}^{* 2}_{N}}
\def\zi{{\zz}^{2r+1}_{2}}
\def\zist{{\zz}^{* 2r+1}_{2}}
\def\ccd{{\cc\hskip0.2truecm}^{2}}
\def\cct{{\cc\hskip0.2truecm}^{3}}
\def\en{{\bf 1 \/}}
\def\pf{{\it pd \/}}
\def\ie{{\it i.e \/}}
\def\tr{{\rm Tr \/}}
\def\eg{{\it e.g \/}} 
\def\cf{{\it c.f \/}} 
\def\viz{{\it viz. \/}}
%Decorated alphabet 
\def\aa {{\cal A \/}}
\def\ad{a^\dag} 
\def\ab{\bar{\alpha}} 
\def\fh{\hat{f}}
\def\nub{\bar{\nu}}
\def\hi{\chi_{klm}} 
\def\udp{U_{\lambda}^{\dagger}}
\def\udm{U_{-\lambda}^{\dagger}} 
\def\utp{\tilde{U}_{\lambda}}
\def\utm{\tilde{U}_{-\lambda}} 
\def\up{U_{\lambda}} 
\def\um{U_{-\lambda}}
\def\td{A^\dag_q}
%Sigmas 
\def\so{\sigma_{1}} 
\def\st{\sigma_{2}} 
\def\sth{\sigma_{3}}
\def\sp{\sigma^{+}} 
\def\sm{\sigma^{-}} 
%Abbreviations
\def\ovl{\overline}

%
%Math I,R,C %\def\IR{\rm I}\!{\rm R}
%\newcommand{\dbl}[2]{\rm#1\hskip-.5em \rm#2} %Bra-kets, absolutes
\newcommand{\bra}[1]{\left<#1\right|} \newcommand{\ket}[1]{\left|#1\right>}
\newcommand{\braket}[1]{\left<#1\right>}
\newcommand{\inner}[2]{\left<#1|#2\right>}
\newcommand{\sand}[3]{\left<#1|#2|#3\right>}
\newcommand{\proj}[2]{\left|#1\left>\right<#2\right|} %
\newcommand{\rbra}[1]{\left(#1\right|} \newcommand{\rket}[1]{\left|#1\right)}
\newcommand{\rbraket}[1]{\left(#1\right)}
\newcommand{\rinner}[2]{\left(#1|#2\right)}
\newcommand{\rsand}[3]{\left(#1|#2|#3\right)}
\newcommand{\rproj}[2]{\left|#1\left)\right(#2\right|}
\newcommand{\absqr}[1]{{\left|#1\right|}^2}
\newcommand{\abs}[1]{\left|#1\right|}
 %Derivatives
\newcommand{\pl}[2]{\partial_{#1}^{#2}} 
\newcommand{\plz}[1]{\partial_{z}^{#1}}
\newcommand{\plzb}[1]{\partial_{\overline{z}}^{#1}} 
\newcommand{\zib}[1]{{\overline{z}}^{#1}} 
%Matrices
\newcommand{\mat}[4]{\left(\begin{array}{cc} #1 & #2 \\ #3 & #4
\end{array}\right)} % 
\newcommand{\col}[2]{\left( \begin{array}{c} #1 \\ #2
\end{array} \right)} 
\newcommand{\qcol}[2]{\left[ \begin{array}{c} #1 \\ #2
\end{array} \right]_q } 
%Gr-alphabet 
\def\a{\alpha} 
\def\b{\beta} 
\def\g{\gamma}
\def\d{\delta} 
\def\e{\epsilon} 
\def\z{\zeta} 
\def\th{\theta} 
\def\f{\phi}
\def\la{\lambda} 
\def\m{\mu} 
\def\p{\pi} 
\def\om{\omega}  
\def\D{\Delta} 
\def\zb{\bar{z}} 
%Others 
\newcommand{\lag}[2]{L_{#1}^{#2}(4\l^{2})}
\newcommand{\mes}[1]{d\mu(#1)} 
%Command abbreviations 
\def\nd{\noindent}
\def\nn{\nonumber} 
\def\cap{\caption} 
\def\cline{\centerline}
\newcommand{\be}{\begin{equation}} 
\newcommand{\ee}{\end{equation}}
\newcommand{\ba}{\begin{array}} 
\newcommand{\ea}{\end{array}}
\newcommand{\bea}{\begin{eqnarray}} 
\newcommand{\eea}{\end{eqnarray}}
\newcommand{\beann}{\begin{eqnarray*}} 
\newcommand{\eeann}{\end{eqnarray*}}
\newcommand{\bfg}{\begin{figure}} 
\newcommand{\efg}{\end{figure}}
%%%%%%%%%%%%%%%%%%%%%%%%%%%%%%%%%%%%%%%%%%%%%%%
\def\ucn{U_{\scriptstyle  CN}}
\def\uca{U_{\scriptstyle  f}}

\title{Quantum Diffusions and Appell Systems
 \footnote{Submitted to Journal of Computational and Applied  Mathematics.
 Special Issue of Proccedings of Fifth Inter. Symp. on Orthogonal Polynomaials, Special Functions and their Applications.
Keywords: Random walks, Hopf Algebras, Appell Systems.}}
\author{Demosthenes Ellinas \footnote{\tt ellinas@science.tuc.gr}
 \\
\\
Department of Sciences\\
Section of Mathematics\\
Technical University of Crete\\
GR - 73 100 Chania Crete Greece}

\maketitle
\begin{abstract}
Within the algebraic framework of Hopf algebras, random walks and associated diffusion
equations (master equations) are constructed and studied for two basic operator algebras of Quantum Mechanics
i.e the Heisenberg-Weyl algebra ($hw$) and its $q$-deformed version $hw_q$. This is done by means of functionals determined by the associated coherent state density operators. The ensuing master equations
admit solutions given by $hw$ and $hw_q$-valued Appell systems.
\end{abstract}

\vskip0.5cm
\nd {\it \bf  1. Introduction. }
 We work in the general framework of the so called quantum probability theory\cite{qprobability} and
 more specifically along the research line  relating random walks,
 diffusions and
 Markov transition operators to Lie-Hopf algebras\cite{mbook,maj1,maj2}. Our aim is to construct algebraic
random walks and their diffusion limit in terms of master equations\cite{risken}. We work with
two basic operator algebras of Quantum Mechanics\cite{gil} i.e the Heisenberg-Weyl algebra ($hw$) and its $q$-deformed version $hw_q$\cite{qosc}, and use their Hopf algebra like structures for
our construction (Chapt. 2). The density of the two functionals needed are constructed by the
associated to those algebras coherent states vectors\cite{klauder}. As the  random
walks take place on the manifold of these coherent states vectors it is important to 
investigate the geometrical features of them (Chapt. 3). Then a limiting
procedure leads to the master (diffusion) equations for the case of $hw$ random
walk (Chapt. 5) and the case of $hw_q$ random walk (Chapt. 6), correspondingly.
The solutions of the resulting master equations of motion for certain
general elements of the respective operator algebras are obtained
in terms of the associated operator valued Appell systems\cite{qappell}. Certain generalities of
classical Appell systems are discussed in Chapt. 4\cite{appell}. Finally, some technicalities
such as ordering formulae for generators of the two $hw$ algebras\cite{katriel}, as well
as some Baker-Campbell-Hausdorff 
  decompositions formulae for the $SU(1,1)$ group elements\cite{wilcox} are summarized in
Appendices A and B.

\vskip0.5cm
\nd {\it \bf  2. Hopf Algebras.}
A Hopf algebra \cite{abe} ${\cal A}={\cal A}(\mu , \eta, \D , \e, S)$ over a
field $k$ is a vector space equipped with an algebra structure
with homomorphic associative product map $\mu:\aa\times\aa\rightarrow\aa$, and a
homomorphic unit map
$\eta:k \rightarrow \aa$, that are related by $\mu\circ(\eta\otimes id)=id=\mu\circ(id\otimes\eta)$, together with a coalgebra structure with a 
homomorphic coassociative coproduct map
$\D:\aa\rightarrow\aa\otimes\aa$ and a homomorphic
counit map $\e:\aa\rightarrow k$, that are related between them by $(\e\otimes id)\circ\D = id = (id\otimes \e)\circ\D $. Both products satisfy
the compatibility condition of bialgebra i.e
$(\mu\otimes\mu)\circ(id\otimes\tau\otimes id)\circ(\D\otimes\D)=
\D\circ\mu $, where $\tau(x\otimes y)=y\otimes x$ stands for the twist map.
%, while the linear antipode map $S:\aa\rightarrow\aa$
%(some kind of generalized inverse) is a antihomomorphsm wrt
%the algebra and coalgebra structures of $\aa$.
If $\eta$ or $\e$ is not defined in $\aa$ we speak about non unital or non counital 
Hopf algebra. 

Suppose we have a functional $\phi:\aa\rightarrow \bf C$, defined on $\aa$, let us
define the operator $T_\phi :\aa \rightarrow \aa$ as $T_\phi =(\phi \otimes id )\circ \D$,
then 
$\e\circ T_\phi = \phi$, namely the counit aids to pass
from the operator to its associated functional.
From this relation we can define the convolution product
$\psi * \phi $, between functionals as follows \cite{maj1}:
\bea
\e\circ T_\psi T_\phi &=& \e \circ (\psi \otimes id )\circ \D \circ
(\phi \otimes id )\circ \D =(\phi\otimes\psi )\circ
(id\otimes id\otimes\e )\circ (id \otimes \D )\circ \D \nn \\
&=&
(\phi\otimes\psi )\circ \D = \phi*\psi \;,
\eea

\nd  and in general
$\e \circ T_{\phi}^n =\e \circ T_{\phi^{*n}}=\phi^{*n}$. These last
relations imply that the transition operators form a discrete
semigroup wrt their composition with identity element $T_\e \equiv id$
(due to the axioms of Hopf algebra) and generator $T_\phi$, while
the functionals form a dual semigroup wrt the convolution with
identity element $e$ and generator $\phi$, and that these two semigroups
are homomorphic to each other.
and the

We recall now two algebras and their structural maps that concerns us
here:\\
i) {\it Heisenberg-Weyl
algebra $hw$}: this is the
algebra of the quantum mechanical oscillator and is generated by
the creation, annihilation and the unit operator $\{ \ad , a, \en \}$ respectively
which
satisfy the commutation relation (Lie bracket) $[a, \ad ]=\en $,
while $\en$ commutes
with the other elements. This algebra possesses a natural non counital
Hopf algebra structure
(or bialagebra-like
cf. \cite{mbook}, Chap. 3), with comultiplication defined
as
\bea 
\D^{(n-1)}a &=& n^{-\frac{1}{2}}(a\otimes\cdots \otimes \en
+ \en\otimes a\otimes \cdots \otimes \en +
\en \otimes \cdots \otimes a)\;, \nn \\
\D^{(n-1)}\ad &=& n^{-\frac{1}{2}}(\ad\otimes\cdots \otimes \en
+ \en \otimes a\otimes \cdots \otimes \en +
\en \otimes \cdots \otimes \ad)\;, \nn \\
\D \en &=& \en \otimes \en + \en \otimes \en \;.
\label{comult}
\eea

\nd Let us also define the so called number operator $N=\ad a$ with
the following commutation relations with the generators of $hw$:

\be
[a,\ad ]=\en \ \ \;, \ [N,\ad ]=\ad  \ \;, \ \ [N,a ]= -a  \;.
\ee

The module which carries the unique irreducible and
infinite dimensional representation of the oscillator algebra is 
the Hilbert-Fock space ${\cal H}_F$ which is generated by a starting
(or "vacuum" ) state vector $\ket{0}\in \cal H$ and is given as
${\cal H}=\{\ket{n}=\frac{(\ad)^{n} }{n!}\ket{0}, n\in {\bf Z}_+$.

ii) {\it The $q$-deformed Heisenberg-Weyl algebra $hw_q$}:  
The $q$-deform Heisenberg-Weyl algebra is generated by
the elements $hw_q= < b, b^\dag , q^N , q^{-N} , \en > $ that satisfy
the relations
\bea
bb^\dag -q^{-1}b^\dag b = q^N \ \ \;, \ \ q^N q^{-N}=\en \;, \nn \\
q^N b q^{-N} =q^{-1}b \ \ \;, \ \ q^N b^\dag q^{-N} =qb^\dag \;.
\eea

\nd For real $q$ the Fock representation space is spanned by the vectors $\{ \ket{n}=
\frac{(b^\dag )^n}{\sqrt{[n]_q !}}\ket{0}, n\in {\bf Z}_+ \}$, where $[n]_q =
 \frac{q^n - q^{-n}}{q-q^{-1}}$
and $[n]_q =[1]_q [2]_q \cdots [n]_q $. In the Fock space representation of this
algebra we have the additional relations $b^\dag b =[N]_q $, $bb^\dag = [N+1]_q $.
This algebras has no satisfactory Hopf structure but still as will be seen below
we can define algebraic random walks on it and study their diffusion limit.
To this end let us make the transformations\cite{qosc} $a_q =q^{N/2 }b$ and $\ad_q =b^\dag q^{N/2 }$,
and obtain the resulting algebra 
\be
a_q \ad_q - q^2 \ad_q a_q =\en 
\label{qqq}
\ee

\nd  which is the new form of the
$hw_q $ algebra\cite{adm}.
Although not an algebra homomorphisms we will use below the coassociative
maps
\bea
\D a_q &=& a_q \otimes \en + \en \otimes a_q \ \ \;,  \ \
\D \ad_q = \ad_q \otimes \en + \en \otimes \ad_q \;.
\label{qcomult}
\eea

\nd {\it \bf  3. Coherent States .}
For our needs here a brief introduction to the concept of
coherent states (CS) on Lie groups
goes as follows: consider a Lie group $\cal G$, with a unitary irreducible 
representation $T(g)$, $g\in \cal G$, in a Hilbert space $\cal H$. We
select a reference vector $\ket{\Psi_0}\in \cal H$, to be called the 
"vacuum" state vector, and let ${\cal G}_0\subset \cal G$ be its isotropy 
subgroup, i.e for $h\in {\cal G}_0$, $T(h)\ket{\Psi_0}=
e^{i\varphi(h)}\ket{\Psi_0}$. The map from the factor group 
${\cal M} = {\cal G}/{\cal G}_{0}$  to the Hilbert space 
$\cal H$, introduced in 
the form of an orbit of the vacuum state under a factor group element,
defines a CSV 
$\ket{x}=T({\cal G}/{\cal G}_{0})\ket{\Psi_0}$ labelled 
by points $x\in\cal M$ of the coherent state manifold.
Coherent states form an (over)complete set of states, since 
by means of the Haar invariant measure of the group $\cal G$ \viz
$\mes{x}, \;\; x\in \cal M$, they provide a resolution of unity,
$\bf{1}=\it{\int_{\cal M}\mes{x}\proj{x}{x}}$. As a consequence, any vector
$\ket{\Psi}\in\cal H$ is analyzed in the CS basis, 
$\ket{\Psi}=\int_{\cal M}\mes{x}\Psi(x)\ket{x}$, with coefficients 
$\Psi(x)=\inner{x}{\Psi}$. We should note here that the square integrability
of the vectors ${\Psi}$ will impose some limits on the growth parameters of
the functions $\Psi(x)$ (cf. \cite{klauder} and references therein).

What concerns us here is mostly the geometry of the CS manifold
$\cal M$. This is due to the fact that the random walks and their diffusion
limits that will be study below will be given in terms of functionals
associated with coherent states so that the random walks will be induced on
the functions defined on $\cal M$ (passive description) or on the operators
acting on the functions defined on $\cal M$ (active description). Although only the
latter description will be studied here in terms of the quantum master equations,
it should be obvious that the geometry of the background manifold $\cal M$ namely
both the Riemannian and the symplectic geometry (the symplectic geometry especially
in the case of non stationary random walks), will manifest itself in
the associated diffusion equations. Specifically below it will be shown that
the $hw$ random walk takes place on the flat complex plane $\bf C$ with canonical
symplectic structure, while the deformed $hw_q$ random walk takes place on a $q$-deformed
surface of revolution with modified, due to $q$-deformation, Riemannian
and symplectic geometry. This fact provides a further motivation for studying 
random walks and diffusions within the present algebraic framework since in this
way we are able to study these phenomena taking place on non trivial spaces. 
Details constructions and studies can be found elsewhere\cite{qcohell}, here we summarize 
some relevant information:

Let us first specialize to the $HW$ group: 
The $hw$-CS is defined by the realation
\be
\ket{\a}=e^{\a\ad-\ab a}\ket{0}=
{\cal N} e^{\a\ad}\ket{0}=e^{-{\fscr{1}{2}}\absqr{\a}}\sum_{n=0}^{\infty}
\frac{\a^n}{\sqrt{n!}}\ket{n}\;.
\ee

\nd It is an (over)complete set of states with respect to the measure
$\mes{\a}=\frac{1}{\pi}e^{-\absqr{\a}} d^{2}\a$ for the non-normalized CS, 
and $\a\in{\cal M}=
HW/U(1)\approx \bf C$ 
is the CS manifold. Since $a\ket{\a}=\a\ket{\a}$,
$\cal M$ is the flat canonical phase plane with the
standard line element $ds^2 =d\a d\ab$. Also the symplectic 2-form 
$\om=id\a \wedge d\ab$ is associated to the canonical Poisson bracket
$\{f,g\}=i(\pl{\a} f \pl{\ab} g - \pl{\ab} f \pl{\a} g) $.

Next we turn to the $hw_q$ case: The definition of the $hw_q$-CS reads\cite{qcohell}
\be
|\ket{\a}_q =e^{\a\ad_q }_q \ket{0}=e^{\a A^\dag_q}\ket{0}=
\sum_{n=0}^{\infty}
\frac{\a^n}{\sqrt{[n]!}}\ket{n}\;,
\ee

\nd where $[n]=\frac{q^{2n}-1}{q^2 -1}$. The states are first defined in terms of the $q$-deformed exponential function $e_q ^x = \sum_{n\geq o} \frac{x^n}{[n]!}$ and the
$q$-creation operator and then equivalently by  exponentiation of the operator $A^\dag_q=\frac{N}{[N]}\ad_q $, that satisfies with
the $hw_q$ elements
the $hw$ algebra relations\cite{qcohell} 
\be
[a_q ,  A^\dag_q ]=\en \ \ \;, \ \ [A_q , a^\dag_q ]=\en \;.
\ee

\nd The $q$-CS is an (over)complete set of states\cite{qcs} with respect to the measure
$\mes{\a}_q =\frac{1}{\pi}(e^{\absqr{\a}}_q )^{-1} d^{2}_q \a$, and wrt the Jackson $q$-integral\cite{ext}. If $q=e^\lambda$, then since $a_q \ket{\a}_q =\a\ket{\a}_q $, the $q$-CS manifold
$\cal M$ is a  non flat  surface of revolution with $q$-deformed induced curvature with curvature scalar
$R=\lambda^2 12(1+2\abs{\a}^2 +{\cal O}(\lambda^3))$. Also the symplectic 2-form $\om$
is modified by the $q$-deformation as $\om=\{i - \frac{\lambda^2}{2}\abs{\a}^2 (\absqr{\a}+2) +{\cal O}(\lambda^3)\}d\a \wedge d\ab$ \cite{qcohell}. 

The density operator (state) $\rho$
will be used below to determine functionals of
some Hopf operator algebras $\cal A$,
so here we introduce the general concept and give its construction
in terms of convex combinations of projectors of coherent states.
Let a Hilbert vector space $\cal H$ that
carries a unitary irreducible representation of $\cal A$ of finite
or infinite dimension. The set
\be
{\cal S}=\{\rho \in {\rm End}({\cal H}): \rho\geq 0 , \rho^\dag =\rho,
tr\rho=1\}\;,
\ee

\nd namely the set of non-negative, Hermitian, trace-one operators acting
on $\cal H$ form a convex subspace of  $ {\rm End}({\cal H})$, 
which is the convex hull of the set
\be
{\cal S}_P = \{\rho \in {\cal S} \ \ , \ \  \rho^2 =\rho \}\equiv {\cal H}/U(1)\;,
\ee

\nd namely of the set of pure density operators (states), that are in
one-to-one correspondance with the state vectors of $\cal H$.
Two kinds of $\rho$ density operators that will be used in the sequel
are constructed by $hw$-CS and $hw_q$-CS. Explicitly
from the pure density operators  $\proj{\pm \a}{\pm \a}\in {\cal S}_P $
and the $q$-deformed ones $\ket{\a}_{qq}\bra{\a}\equiv \proj{\a}{\a}_q \in {\cal S}_P $,
we form convex combination belonging to the convex hull of ${\cal S}_P$
i.e
\bea
\rho &=& p\proj{\a}{\a} +(1-p)\proj{-\a}{-\a}\;, \nn \\
\rho_q &=& p\proj{\a}{\a}_q +(1-p)\proj{-\a}{-\a}_q\;.
\eea

\vskip0.5cm
\nd {\it \bf 4. Appell Systems.}
Classical Appell polynomials \cite{appell} on the real line are  polynomials
$\{ h_n (x); n\in \bf N\}$ of degree $n$ that satisfy the condition
$(d/dx) h_n (x)=nh_n (x) $. A class of such systems is the
shifted moment sequences $h_n (x)=\int_{-\infty}^{\infty}
(x+y)^n p(dy)$, for some positive real measure $p$ with finite moments.
The class of Appell polynomials includes cases such as the divided
sequences, the Bernoulli polynomials and the Hermite polynomials,
which correspond to the Gaussian measure $p=p(dy)=\frac{1}{\sqrt{2\pi}}
e^{-y^2 /2}dy$. Some important properties of the Appell polynomial sets that have been investigated are the following: Hermite polynomials
are the only Appell polynomials associated to the ordinary derivative
operator that are also orthogonal
\cite{appell}(e, b, c), similarly Charlier polynomials are
the only Appell systems associated to the difference operator that are
also orthogonal\cite{appell}(d), while the Rogergs $q$-Hermite polynomials are
the only Appell systems associated to Askey-Wilson $q$-derivative
operator that are orthogonal too\cite{appell}(e).

The following Hopf algebraic reformulation of the
real line Appell systems (i.e non polynomials necesserily)
motivates their generalization to more general 
spaces. Let ${\cal A}={\bf R}[[X]]$ the algebra
of the real formal power series generated by pointwise multiplication
$fg(x)=f(x)g(x), f,g \in \cal A$. Then ${\cal A}$ becomes a Hopf
algebra with comultiplication $(\D f) (x,y)=f(x+y)$ and counit $\e (id)=1$,
$\e(X)=0$, where $id$ is the identity function and $X(x)=x$ stands
for the coordinate function. For a given functional $\phi:{\cal A}
\rightarrow \bf C$ and a chosen basis $( x^n ), n\in {\bf Z}_+ $ in
$\cal A$, it is easy to verify that the relation
$h_n (x)=(\phi \otimes id)\circ\D x^n = T_\phi x^n$ defines an Appell systems
and is equivalent to the preceding
definition. Specifically for $\phi=\int_{-\infty}^{\infty} p(dy)$
with $p$ the Gaussian measure we obtain the Hermite polynomials if
we make the identifications $x\otimes 1 \equiv x$ and $1\otimes x \equiv iy$.
This algebraic definition has been used extensively to introduce
Appell systems in non commuting algebras\cite{qappell}.
Here we will utilize it to define below Appell systems on two important
operator algebras of Quantum Mechanics i.e the Heisenberg algebra
and the $q$-deformed Heisenberg algebra and to show that the resulting 
operator valued Appell systems are 
solutions of quantum master equations that are constructed
respectively
as limits of random walks defined on these algebras.

\vskip0.5cm
\nd {\it \bf  5. Diffusion on C.}
Let $\phi(\cdot)={\rm Tr}\rho(\cdot)\equiv<\rho ,\cdot >$, a
functional defined on the enveloping Heisenberg-Weyl algebra
${\cal U}(hw)$, where
$\rho=p\proj{\a}{\a}+(1-p)\proj{-\a}{-\a}$, i.e the $\rho$ density
operator is given as a convex sum of pure state density operators.
The action of the transition operator $T_\phi = (\phi \otimes
id)\circ \D $ on the generating monomials of ${\cal U}(hw)$ (where we ignore the numerical
factors in the comultiplication of eq.(\ref{comult})) reads,
\bea
T_\phi ((\ad)^{m} a^{n} ) &=& (\phi \otimes id)\circ \D ( (\ad)^{m} a^{n} )\nn \\
&=& \sum_{i=0}^{m} \sum_{j=0}^n \col{m}{i}\col{n}{j}
[p \a^{* i}\a^{j} + (1-p)(-\a)^{i} (-\a)^j ] (\ad)^{m-i}a^{n-j}\nn \\
&=& p (\ad + \a^{*} )^{m} (a + \a )^{n} + (1-p) (\ad - \a^{*} )^m
(a- \a )^n \;.
\eea

\nd For a general element $f(a,\ad ) \in {\cal U}(hw)$ that is normally ordered, namely the
annihilation operator $a$ is placed to the right of
the creation operator $\ad$, denoted by
$\fh (a,\ad ) =\sum_{m,n \geq 0}c_{mn} (\ad)^m a^n$, the action of
the linear operator $T_\phi$ becomes
\be
T_\phi  (\fh(a,\ad ) ) = p \fh(a + \a ,\ad +\a^* ) +
(1-p) \fh(a - \a ,\ad -\a^* )
\ee

%\nd The counit map is defined by $\e (a^j )=\e(\ad^i)=0 $ for $i,j \geq 1$
%and $\e({\bf 1})=1$, then due to its homomorphism 
%we have that

\nd By means of the CS eigenvector property and the normal ordering
of the $f$ element we also compute the value of functional viz.
\be
\phi (\fh(a,\ad ))= p \fh( \a ,\a^* ) +
(1-p) \fh( - \a , -\a^* ) \;.
\ee

Let us consider the displacement operator $D_\a
= e^{\a \ad - \a^* a}$ which acts with the group adjoint action
on any element $f$ of the ${\cal U}(hw)$ algebra viz.\cite{klauder}
\be
Ad D_a (f)=Ad e^{\a \ad - \a^* a}(f)=
Ad e^{ad (\a \ad - \a^* a)}(f)=  D_\a f D_\a ^\dagger ,
\ee

\nd where $ad (X)f=[X,f]$ and $ad(X)ad(X)f =[X,[X,f]]$ and
similarly for higher powers, stands for the Lie algebra
adjoint action that is defined in terms of the Lie commutator.
Explicitly the action of the displacement operator on the generators
of ${\cal U}(hw)$ reads $Ad D_{\pm \a}(a)=a\mp \a$ and
$Ad D_{\pm \a}(\ad )=\ad \mp \a^* $. By means of these expressions
we rewrite the action of the preceding transition operator
as 
\be
 T_\phi (\fh(a,\ad )) =[p Ad D_{-\a} + (1-p) Ad D_{\a}]\fh
%=pD_\a^\dagger \fh D_\a +(1-p)
%D_{-\a}^\dagger \fh D_{-\a}.
\ee

\nd Next we want to compute the limiting transition operator
\bea
T_t &\equiv &  T_{\phi_t} \equiv 
\lim_{n\rightarrow \infty}T_\phi^n \nn \\
&=&\lim_{n\rightarrow \infty} [p (1+ ad (-\a \ad + \a^* a) +\frac{1}{2}ad ad (-\a \ad + \a^* a)+\cdots )\nn \\
&+&(1-p)(1+ ad (\a \ad - \a^* a) +\frac{1}{2}ad ad (\a \ad - \a^* a)+\cdots ]^n
\;.
\eea

\nd If we introduce the parameters $t\in{\bf R}$ and $c,\gamma \in {\bf C}$
by means of the relations,
\be
2\a(p-\frac{1}{2} )=\frac{tc}{n}  \ \ \ , \ \ \ \frac{\a^2}{2}=
\frac{t\gamma}{n}\;,
\label{relations}
\ee

\nd and then take $\a\rightarrow 0, n\rightarrow \infty$, with $t, c, \gamma$ fixed, 
we use the limit $(1+\frac{Z}{n})^n \rightarrow e^Z$, to
arrive at the limiting Markov operator $T_t =e^{t ad {\cal L}}$,
where 
\be
{\cal L}=-c\ad +c^{*} a +\gamma (\ad)^{2} -\gamma^{*} a^{2} - \abs{\gamma}
(\ad a +a\ad ).
\ee
 
By construction $T_t$ is the time evolution operator
for any element  $f$ of ${\cal U}(hw)$ i.e $f_t =T_t (f)$ and
forms a continous semigroup $T_t T_{t'} = T_{t+t'}$ under
composition. This yields the
diffusion
equation obeyed by  $f_t$, which will be taken to
be normally ordered hereafter. By time derivation of the equation
\be
\phi_t (\fh)=<\rho , \fh_t>=
<\rho , e^{t ad {\cal L}}\fh>=< e^{-t ad{\cal L}^\dagger }\rho , \fh>
=<\rho_t , \fh>\;,
\ee

\nd we obtain the
diffusion equation $\frac{d}{dt}\fh_t  ={\cal L}\fh_t  $, as well as the dual
one satisfied by the $\rho$ density operator viz. $\frac{d}{dt}\rho_t =
{\cal L}^\dagger \rho_t $.
To simplify and eventually solve the ensuing equations we will
assume here that the
parameter $\gamma$ introduced above is a complex variable with
random argument of zero
average and constant non zero magnitude. Then if we average over random
$\gamma$  the equations of motion only the term proportional to the
amplitude of $\gamma$ will be retained. If in addition we consider the
case of an symmetric random walk i.e $p=1/2, c=0$ the equation
of motion becomes
\be
\frac{d}{dt}\fh_t   =-2\abs{\gamma} \left[ \ad \fh_t a
+a \fh_t \ad - N\fh -\fh(N+1)\right]\;.
\label{master}
\ee

\nd This is a quantum master equation of the Lindblad type\cite{lind} which will be
shown to admit a solution in terms of a operator valued Appell system
associated with the generator of that equation. We may
introduce the following operators\cite{davies}
\be
K_+ f=\ad f a \ \, \ \
K_- f = af \ad \ \, \ \ K_0 f =\frac{1}{2}(\ad a f + f a\ad )\;,
\label{oper}
\ee

\nd and $K_c f =[\ad a , f ]$. These operators acting on the elements $f$
of the enveloping algebra ${\cal U}(hw)$, generate the
$su(1,1)$ Lie algebra  defined by the commutation relations
\be
[K_- , K_+ ]=2K_0 \ \ \  ,\ \ \ [K_0 , K_\pm ]=\pm K_\pm ,
\label{comrel}
\ee

\nd where $K_c $ is the central element (Casimir operator ) of the algebra.
In terms of these operators the quantum master equation (\ref{master})
 is cast in the form 
\be
\frac{d}{dt}\fh_t  =-2\abs{\gamma}(-2K_0 + K_+ + K_- )
 \fh_t \;.
\ee

\nd  Use of the disentangling theorem (Baker-Campbell-Hausdorff formula)
 of a general $SU(1,1)$ group element (c.f Appendix A),
allows  to express the solution of the quantum master equation in the
form 
\be
\fh_t = \exp(A_+ K_+ )\exp(\ln A_0 K_0 ) \exp(A_- K_- )(\fh)
=\exp(B_- K_- )\exp(\ln B_0 K_0 )
 \exp(B_+ K_+ )(\fh)\;,
\ee

\nd if the  normally or respectively antinormally ordered BCH decomposition
is used.
Above $\fh=\sum_{n\geq 0}c_{mn}(\ad)^{s} a^{t} $, stands for the initial time
operator which can be a general element of the enveloping algebra
${\cal U}(hw)$.
Specifically in the case of normally ordered decomposition with
initial operator taken as $\fh= (\ad)^{m} a^{n}$  the solution of the
quantum master equation is obtained by means of the actions issued in
eq.(\ref{oper})
and by the antinormal-to-normal reordering relations among the generators of
the  ${\cal U}(hw)$ algebra (c.f Appemdix B). An arduous but straightforward
calculation yields the normal ordered solution:
\bea
 \hspace*{-6.0cm}\fh_t  = \exp(A_+ K_+ )
 \exp(\ln A_0 K_0 ) \exp(A_- K_- )((\ad)^{s} a^{t} ) =
 \nn
\eea
\bea
   \hspace*{-3.70cm}\sum_{k\geq 0}\sum_{l\geq 0}\sum_{m\geq 0}
\sum_{i=0}^{min(k,s)}
\sum_{j=0}^{min(k+t-i,k)}
\sum_{u=0}^{l}
\sum_{v=0}^{u}
\sum_{q=0}^{l-u}
\sum_{w=0}^{v}
\sum_{f=0}^{min(q,x)}
\sum_{h=0}^{min(y+q-f,w)} \times
\nn
\eea
\bea
 \frac{A_- ^k}{k!}\frac{\overline{A}_0 ^l}{l!}\frac{A_+ ^m}{m!}
d_{k,s}^{i}d_{k+t-i,k}^{j}d_{q,x}^{f}d_{y+q-f,w}^{h}\overline{d}_{l-u,q}
\overline{d}_{v,w } \frac{1}{2^l } \col{l}{u}\col{u}{v} (\ad)^{x+w+m-f-h}a^{y+q+m-f-h}
\label{solution}
\eea

\nd where $x=s+k+q-i-j$, $y=t+k+w-i-j$ and $\overline{A}_0 = \ln A_0$, with
$A_0 =\frac{1}{1-4\abs{\gamma}t}$ and $A_\pm = \frac{-2\abs{\gamma}t}{1-2\abs{\gamma}t}$.
 A similar solution can be
obtained for the antinormal BCH decomposition. We can therefore state
the results in the following

{\bf Proposition 1.}
{\it The solution of the quantum master equation
$\frac{d}{dt}\fh_t  ={\cal L}\fh_t  $
where the generator ${\cal L}(\fh_t)
=-2\abs{\gamma} \left[ \ad \fh_t a
+a \fh_t \ad - N\fh -\fh(N+1)\right]$
of Lindblad type
generates the semigroup of Markov transition operators $T_t = e^{t{\cal L}}$
%the semigroup of functionals $\phi_t =\e \circ e^{t{\cal L}}$
acting on the enveloping algebra
${\cal U}(hw)$, is given by the associated ${\cal U}(hw)$-valued
Appell system which in its
normally ordered form is given by equation (\ref{solution})}.

 We note also that the dual master equation satisfied by the density
operator can easily be solved along the above lines in terms of
the associated Appell system.

\vskip0.5cm
\nd {\it \bf 6. $q$-Diffusion.}
Let $\phi_\phi (\cdot)={\rm Tr}\rho_q(\cdot)\equiv<\rho_q ,\cdot >$, a
functional defined on the enveloping $q$-Heisenberg-Weyl algebra
${\cal U}_q (hw)$, where
$\rho_q =p\proj{\a}{\a}_q +(1-p)\proj{-\a}{-\a}_q$ is the $\rho$ density
operator given as a convex sum of pure state $q$-density operators.
The action of transition operator $T_\phi^q = (\phi_q \otimes
id)\circ \D $ on the monomials of ${\cal U}_q (hw)$, with $\D$ map given is
eq.(\ref{qcomult}) reads,
\bea
T_{\phi_q } (\ad)^m_q a^n_q )&=& (\phi_q \otimes id)\circ \D ( (\ad)^m_q a^n_q )\nn \\
&=& \sum_{i=0}^{m}\sum_{j=0}^n \col{m}{i}\col{n}{j}
[p \a^{* i} \a^j + (1-p)(-\a)^{i} (-\a)^{j} ] (\ad)^{m-i}_q a^{n-j}_q \nn \\
&=& p (\ad_q + \a^* )^{m} (a_q + \a )^{n} + (1-p) (\ad_q - \a^{*} )^{m}
(a_q - \a )^{n} \;.
\eea

\nd On an element $f(a_q ,\ad_q )$ of the enveloping algebra
${\cal U}_q(hw)$ that is normally ordered, namely the
annihilation operator $a_q$ is placed to the right of
the creation operator $\ad_q$, that is expressed as 
$\fh(a_q ,\ad_q ) =\sum_{m,n \geq 0}c_{mn} (\ad)^{m}_{q} a^{n}_{q}$, the action of
the linear operator $T_{\phi_q}$ becomes
\be
T_{\phi_q}(\fh(a_q ,\ad_q )) = p \fh(a_q + \a ,\ad_q +\a^* ) +
(1-p) \fh(a_q - \a ,\ad_q -\a^* )\;.
\ee

%The counit is defined $\e (a^j_q )=\e(\ad^i_q)=0 $ for $i,j \leq 1$
%and $\e({\bf 1})=1$, then due to homomorphism of counit
%we have that

By means of the $q$-CS eigenvector property and the normal ordering
of the  element $f$ we also compute the value of functional viz.
\be
%\e \circ T_{\phi_q}( \fh(a_q ,\ad_q ))
\phi_q \fh(a_q ,\ad_q )= p \fh( \a_q ,\a^* _q ) +
(1-p) \fh( - \a , -\a^* ) \;.
\ee

Let us now consider the displacement operator $D_\a ^q
= e^{\a \td - \a^* a_q }$, which acts with the following adjoint action
on any element $f$ of the ${\cal U}_q(hw)$ algebra,
$Ad D_a ^q (f)=Ad e^{\a \td - \a^* a_q }(f)=
e^{ad (\a \td - \a^* a_q )}(f)=  D_\a ^q f D_{-\a} ^{q } $. We should emphasize at
this point that $D_\a ^{q \dag}\neq D_{-\a}^q $. This is an important
difference from
the preceding undeformed case with $q=1$, which stems from the fact the though eq.(\ref{qqq}) is valid the two involved operators are not Hermitian conjugate to each other. This
fact  would not permit us to proceed for
the construction of quantum diffusion equation in a manner analogous to the $q=1$ case.
Instead here we will restrict the space of solutions
of the resulting $q$-master equation from
the whole algebra ${\cal U}_q (hw)$ to the commuting subalgebra
generated either
by monomials of the creation operator $\{ (\ad)^{m}_{q} , m\in {\bf Z}_+ \}$ or
of the annihilation operator
$\{a^m_q , m\in {\bf Z}_+ \}$ alone. Notice however that such a choice would be undesirable from the physical point
of view since it would not allow us to study Hermitian solutions of the ensuing 
master equation.

\nd Then the explicit action of the $q$-displacement operator on the generators
of ${\cal U}_q (hw)$  reads $Ad D_{\pm \a}^q (a_q )=a_q \mp \a$ and
$Ad D_{\mp \a}^{q \dag}(\ad_q )=\ad_q \mp \a^* $. By means of these expressions
we rewrite the action of the preceding $q$-transition operator on an
analytic formal power series $f(a_q )$
as 
\be
 T_{\phi_q}( f(a_q )) =[p Ad D_{-\a} ^q + (1-p) Ad D_{\a} ^q ]
(f(a_q ))\;.
\ee

We wish to compute the limiting transition operator
\bea
T_t ^q &\equiv & T_{\phi_t^q }\equiv \lim_{n\rightarrow\infty}
(T_{\phi_q})^n \nn \\
%\lim_{n\rightarrow \infty}T_\phi^{q n} \nn \\
&=&\lim_{n\rightarrow \infty} [p (1+ ad (-\a \td + \a^* a_q )
+\frac{1}{2}ad ad (-\a \td + \a^* a_q )+\cdots ) \nn \\
&+& (1-p)(1+ ad (\a \td - \a^* a_q )+ \frac{1}{2}ad ad (\a \td - \a^* a_q)
+\cdots ]^n
\;.
\eea

\nd If we introduce the parameters $t\in{\bf R}$ and $c,\gamma \in {\bf C}$
by means of the same relations (\ref{relations}) as in the $q=1$ case,
then we will obtain the limiting $q$-transition operator
$T_t ^q =e^{t ad {\cal L}_q }$,
where ${\cal L}_q = - c\td + c^{*} a_{q} +\gamma (\td)^2 -\gamma^{*} 
a^{2}_{q} -
\abs{\gamma}
(\td a_q +a_q \td )$.

To simplify this $q$-master equation we will
assume as in the undeformed case that the
parameter $\gamma$ is a complex variable with
random argument of zero
average and constant non zero magnitude. Then if we average over random
$\gamma$  the equation of motion then only terms proportional to the
amplitude of $\gamma$ will be retained. If in addition we consider the
case of an symmetric random walk i.e $p=1/2, c=0$ the equation
of motion becomes
\be
\frac{d}{dt}f_t   =-2\abs{\gamma} \left[ \td f_t a
+a f_t \td - Nf_t -f_t(N+1)\right]\;.
\label{qmaster}
\ee

This is a $q$-quantum master equation of the Lindblad type\cite{lind} which will be
shown to admit a solution in terms of a operator valued Appell system
associated with the generator of that equation. We may
introduce as in the preceding undeformed case the following operators
\be
K_+ f=\td f a_q \ \, \ \
K_- f = a_q f \td \ \, \ \ K_0 f =\frac{1}{2}(\td a_q f + f a_q \td )\;,
\label{qoper}
\ee

\nd and $K_c f =[\td a_q , f ]$. These operators acting on the elements $f$
of the enveloping algebra ${\cal U}_q (hw)$, generate the
$su(1,1)$ Lie algebra  defined as in eq. (\ref{comrel}).
In terms of these operators the $q$-quantum master equation (\ref{qmaster})
 is cast in the form 
\be
\frac{d}{dt}f_t  =-2\abs{\gamma}(-2K_0 + K_+ + K_- )
 f_t \;.
\ee

\nd  Use of the disentangling theorem (Baker-Campbell-Hausdorff formula)
 of a general $SU(1,1)$ group element (c.f Appendix A),
allows  to express the solution of the quantum $q$-master equation in the
form 
\be
\fh_t = \exp(A_+ K_+ )\exp(\ln A_0 K_0 ) \exp(A_- K_- )(\fh)
=\exp(B_- K_- )\exp(\ln B_0 K_0 )
 \exp(B_+ K_+ )(\fh)\;,
\ee

\nd if the  normally or respectively the antinormally ordered BCH decomposition
is used.
Above we choose $f=\sum_{n\geq 0}c_{n} a^n_q $, to stand for the initial time
operator which can be a general element of the subalgebra of
${\cal U}_q (hw)$ that is generated by the $q$-annihilation operator.
Specifically in the case of normally ordered decomposition with
initial operator taken as $f= a_q^t$  the solution of the
quantum $q$-master equation is obtained by means of the actions issued in
eq.(\ref{qoper}).
A straightforward
calculation yields the solution:
\bea
 \hspace*{-4.0cm}f_t  = \exp(A_+ K_+ )\exp(\ln A_0 K_0 )
 \exp(A_- K_- )(a^t ) =
 %\nn
%\eea
%\bea
%\hspace*{-3.70cm}
\sum_{k\geq 0}\sum_{l\geq 0}\sum_{m\geq 0}
\sum_{r=0}^{l}\times \nn \\
%\nn
%\eea
%\bea
 \frac{A_- ^k}{k!}\frac{\overline{A}_0 ^l}{l!}\frac{A_{+}^{m}}{m!}
\frac{1}{2^{m}}\col{l}{r}
(\ad)_q^{k+m}(N+k)^{l-r}(N+k+t+1)^{r}a_q^{k+t+m}\;,
\label{qsolution}
\eea

\nd where the $A$'s have the same values as before.
% as before $\overline{A}_0 = \ln A_0$, with
%$A_0 =\frac{1}{1-4\abs{\gamma}t}$ and $A_\pm = \frac{-2\abs{\gamma}t}
%{1-2\abs{\gamma}}$.
 A similar solution can be
obtained for the antinormal BCH decomposition. We can therefore state
the results in the following

{\bf Proposition 2.}
{\it The solution of the quantum $q$-master equation
$\frac{d}{dt}f_t  ={\cal L}_q f_t  $
where the operator ${\cal L}_q (f_t)
=-2\abs{\gamma} \left[ \td f_t a_q
+a_q f_t \td - Nf_t -f_t (N+1)\right]$
of Lindblad type
generates the semigroup of $q$-Markov transition operators $T_t^q = e^{t{\cal L}_q}$
acting on the enveloping algebra
${\cal U}_q(hw)$, is given by the associated $\ad_q a_q$-valued
Appell system which 
is given by equation (\ref{qsolution})}.

We note also that the dual $q$-master equation satisfied by the density
operator can easily be solved along the above lines in terms of
the associated Appell system.

\vskip0.5cm
\nd {\bf 7. Discussion}. A novel way for constructing quantum master equations
has been provided with solutions given by certain sets of operator valued
functions that constitute a generalization of the concept of classical
Appell polynomial. This entire approach is algebraic and utilizes
concepts and tools from the powerfully structure of Hopf algebra. 
The choice of the dual partner of that algebra structure, namely the coherent states  and their adjoint
density operators, offers a chance to investigate random walks on non trivial geometries.

The prospect of such a framework is rich enough to allow for random walks constructed
on e.g non commuting spaces with braided/smash structure\cite{elltso} or on Lie groups,
quantum groups and quantum modules and comodules. The kinds of Appell systems resulting 
in those cases might provide new challences to the theory of
Special Functions. Some of these issues will be taken up in a forthcoming 
communication\cite{ell}.

\vskip0.5cm
\nd {\it \bf Appendix A.}
The disentangling theorem (Baker-Campbell-Hausdorff formula)\cite{gil}
 of a general  $SU(1,1)$ group element\cite{wilcox}   
$g(a_+ , a_0 , a_- )$
 in the normal 
 $\{ K_+ ^a K_0 ^b K_- ^c : a, b, c \in {\bf Z}_+ \}$, and antinormal
 $\{ K_- ^a K_0 ^b K_+ ^c  : a, b, c \in {\bf Z}_{+} \}$ ordering of the
 generators of the enveloping algebra ${\cal U}(su(1,1))$
 reads respectively:
\bea
g(a_+ , a_0 , a_- )&=& \exp(\a_+ K_+ + \a_0 K_0 +\a_- K_- )\nn \\
&=&\exp(A_+ K_+ )\exp(\ln A_0 K_0 )
 \exp(A_- K_- )\;, \nn \\
&=&\exp(B_- K_- )\exp(\ln B_0 K_0 )
 \exp(B_+ K_+ )\;,
 \eea

\nd where $A_\pm(a_0)=\frac{(a_\pm / \phi)\sinh\phi}{\cosh\phi-
(a_0 /2\phi )\sinh\phi }$, $A_0 =(\cosh\phi -(a_0 /2\phi )\sinh\phi)^{-2}$
and $B_\pm (a_0)=-A_\pm (-a_0)$, $B_0 =(\cosh\phi +(a_0 /2\phi)
\sinh\phi )^2 $, with $\phi^2 =((\a_0 /2 )^2 -a_+ a_- )$. The relations
between the two types of ordered decompositions is based on the formulae
$A_\pm = \frac{B_0 B_\pm }{1-B_0 B_+ B_- }$, $A_0 =\frac{B_0 }
{(1-B_0 B_+ B_- )^2 }$, and $B_\pm =\frac{A_\pm}{A_0 - A_+ A_- }$,
$B_0 = 1/A_0 (A_0 - A_+ A_- )^2 $.

\vskip0.5cm
\nd {\it \bf Appendix B.}
Relations among ordered basic monomials of the enveloping algebra
${\cal U}(hw)$\cite{katriel}.
From antinormal to normal ordering:
\be
a^i (\ad)^j =\sum_{l=0}^{min(i,j)}d^l _{i,j}(\ad)^{j-l}a^{j-l}
=\sum_{l=0}^{min(i,j)} l!\col{i}{l}\col{j}{l}(\ad)^{j-l}a^{j-l}
\;,
\ee

\nd From number operator to normal ordering:
\be
N^k =\sum_{l=1}^k c_{k,l}(\ad)^l a^l \;,
\ee
\nd where $\overline{c}_{k+1,l}=\overline{c}_{k,l-1}+l\overline{c}_{k,l}$,
and these coefficients are recognized as the Stirling numbers of
second kind.

\nd From number operator to antinormal ordering:
\be
N^k =\sum_{l=1}^k \overline{d}_{k,l}a^l (\ad)^l
 \;,
\ee

\nd where $\overline{d}_{k+1,l}=\overline{d}_{k,l-1}-(l+1)\overline{d}_{k,l}$,
with $\overline{d}_{0,0}=1$.

Relations among ordered basic monomials of the enveloping algebra
${\cal U}_q (hw)$\cite{katriel}.
From antinormal to normal ordering:
\be
a^i_q (\ad)^j_q =\sum_{l=0}^{min(i,j)}\overline{b}^l _{i,j}
(\ad)^{j-l}_q a^{j-l}_q =\sum_{l=0}^{min(i,j)}q^{l(l-i-j)+ij}[l]!
\qcol{i}{l}\qcol{j}{l}(\ad)^{j-l}_q a^{j-l}_q
\;.
\ee

\nd We note that for $q\rightarrow 1$ the
$\overline{b}^{l}_{i,j}\rightarrow d^{l}_{i,j}$. From normal to antinormal
ordering:
\be
(\ad)^{i} a^{j}=\sum_{l=0}^{min(i,j)}b^{l}_{i,j} a^{j-l}(\ad)^{i-l}
=\sum_{l=0}^{min(i,j)}(-)^{l} q^{l(l-i-j)-ij)} [l]! \qcol{i}{l}\qcol{j}{l}
a^{j-l}(\ad)^{i-l}
\;.
\ee

\vskip0.5cm
\nd {\bf Acknowledgement}. Discussions with I. Tsohantjis are gratefully acknowledged.

\end{document}